\documentclass[aps,
reprint,
prl, 
amsmath,
amsfonts,
amssymb,
superscriptaddress,
]{revtex4-2}

\usepackage[
breaklinks=true,
colorlinks=true,
bookmarksopen,
bookmarksnumbered,
linkcolor=teal,
filecolor=DarkBlue,
urlcolor=PRDblue,
citecolor=tclr,
pdftex
]{hyperref}
\usepackage[numbered]{bookmark}

\usepackage{tikz}
\usetikzlibrary{automata,positioning}
\usetikzlibrary{snakes}
\usepackage{fontawesome} 

\newcommand{\beq}{\begin{equation}}
\newcommand{\eeq}{\end{equation}}
\newcommand{\bea}{\begin{eqnarray}}
\newcommand{\eea}{\end{eqnarray}}
\newcommand{\be}{\begin{equation}} 
\newcommand{\ee}{\end{equation}}
\newcommand{\ba}{\begin{eqnarray}}
\newcommand{\ea}{\end{eqnarray}}
\newcommand{\alp}{\alpha}

\newcommand{\lam}{\lambda}

\newcommand{\Leff}{\Lambda_{\rm eff}}

\newcommand{\newc}{\newcommand}
\newc{\pa}{\partial}
\def\eq$#1${\begin{equation}#1\end{equation}}
\def\gat$#1${\begin{gather}#1\end{gather}}
\def\bal$#1${\begin{align}#1\end{align}}
\newc{\nonum}{\nonumber}

\usepackage{times}
\usepackage[top=1in, bottom=0.4in, left=0.55in, right=0.55in,includefoot]{geometry}
\definecolor{DarkViolet}{RGB}{148,0,211}
\definecolor{brightcerulean}{rgb}{0.11, 0.67, 0.84}
\definecolor{babyblue}{rgb}{0.54, 0.81, 0.94}
\definecolor{nicegreen}{RGB}{0,153,0}
\definecolor{nicegred}{RGB}{255,66,66}
\definecolor{coral}{rgb}{1.0, 0.5, 0.31}
\definecolor{azure(colorwheel)}{rgb}{0.0, 0.5, 1.0}
\definecolor{cerulean}{rgb}{0.0, 0.48, 0.65}
\definecolor{nicered}{rgb}{0.7,0.1,0.1}
\definecolor{DarkBlue}{rgb}{0,0.2,0.65}
\definecolor{PRDblue}{RGB}{48,46,146}
\definecolor{tikz-orange}{RGB}{255,216,178}
\definecolor{deeplilac}{rgb}{0.7, 0.13, 0.83}

\definecolor{tclr}{RGB}{148,0,211}
\newcommand{\SectionColor}[1]{\textcolor{PRDblue}{#1}}
\newcommand{\SectionStyle}[1]{\SectionColor{\textit{\textbf{#1}}}}
\usepackage{scalerel}
\usetikzlibrary{svg.path}
\definecolor{orcidlogocol}{HTML}{A6CE39}
\tikzset{orcidlogo/.pic={\fill[orcidlogocol] svg{M256,128c0,70.7-57.3,128-28,128C57.3,256,0,198.7,0,128C0,57.3,57.3,0,128,0C198.7,0,256,57.3,256,128z};\fill[white]svg{M86.3,186.2H70.9V79.1h15.4v48.4V186.2z}svg{M108.9,79.1h41.6c39.6,0,57,28.3,57,53.6c0,27.5-21.5,53.6-56.8,53.6h-41.8V79.1z M124.3,172.4h24.5c34.9,0,42.9-26.5,42.9-39.7c0-21.5-13.7-39.7-43.7-39.7h-23.7V172.4z}svg{M88.7,56.8c0,5.5-4.5,10.1-10.1,10.1c-5.6,0-10.1-4.6-10.1-10.1c0-5.6,4.5-10.1,10.1-10.1C84.2,46.7,88.7,51.3,88.7,56.8z};}}
\newcommand\orcid[1]{\href{https://orcid.org/#1}{\mbox{\scalerel*{
\begin{tikzpicture}[yscale=-1,transform shape]
\pic{orcidlogo};
\end{tikzpicture}
}{|}}}}

\makeatletter
\newcommand*{\sectionbookmark}[1][]{%
  \bookmark[%
    level=section,%
    dest=\@currentHref,%
    #1%
  ]%
}
\makeatother

\begin{document}
\title{\Large{Bridging Dimensions: General Embedding Algorithm and \\ Field-Theory Reconstruction in 5D Braneworld Models}}

\author{Theodoros Nakas\,\orcid{0000-0002-3522-5803}}
\email[]{theodoros.nakas@gmail.com}
\affiliation{Physics Division, School of Applied Mathematical and Physical Sciences,
National Technical University of Athens, Zografou Campus, Zografou, GR-15780, Greece}
\affiliation{Research Centre for Theoretical Physics and Astrophysics, Institute of Physics, Silesian University in Opava, Bezručovo náměstí 13, CZ-746 01 Opava, Czech Republic}
\author{Thomas D.~Pappas\,\orcid{0000-0003-2186-357X}}
\email[]{thomas.pappas@physics.slu.cz}
\affiliation{Research Centre for Theoretical Physics and Astrophysics, Institute of Physics, Silesian University in Opava, Bezručovo náměstí 13, CZ-746 01 Opava, Czech Republic}
\author{Zden\v{e}k Stuchl\'ik}
\email[]{zdenek.stuchlik@physics.slu.cz}
\affiliation{Research Centre for Theoretical Physics and Astrophysics, Institute of Physics, Silesian University in Opava, Bezručovo náměstí 13, CZ-746 01 Opava, Czech Republic}

\begin{abstract}
\noindent\textbf{Abstract.}
We develop a general algorithm that enables the consistent embedding of any four-dimensional static and spherically symmetric geometry into any five-dimensional single-brane braneworld model, characterized by an injective and nonsingular warp factor. 
Furthermore, we supplement the algorithm by introducing a method that allows one to, in principle, reconstruct 5D field theories that support the aforementioned geometries.
This approach is based on a conformal transformation of the metric with the conformal factor being identified with the warp factor of the bulk geometry. 
The reconstructed theories depend solely on the induced brane geometry, since the warp factor is model-independently represented by a scalar field in the Lagrangian density. 
As a first application of our reconstruction method, we present for the first time a complete theory that supports the five-dimensional brane-localized extension of the Schwarzschild black hole, for any warp factor.
The same method is subsequently utilized to illustrate the process of coherently embedding a de Sitter brane in braneworld models. 
\end{abstract}

\maketitle


\SectionStyle{Introduction}\sectionbookmark{Introduction}---The Randall-Sundrum model~\cite{Randall:1999ee,Randall:1999vf}, initially proposed as a solution to the hierarchy problem of particle physics, made its appearance in the late 1990s, and has thereafter revolutionized the way we think about extra dimensions.
It has also served as a cornerstone for the development of subsequent models with novel physical characteristics~\cite{Csaki:1999jh, Csaki:1999mp, Goldberger:1999uk, Verlinde:1999fy, Pomarol:1999ad, Davoudiasl:1999tf, Giddings:2000mu, Davoudiasl:2000wi, Huber:2000ie, Agashe:2002pr, Contino:2003ve, Agashe:2003zs}.
In the context of braneworld models, our four-dimensional (4D) Universe, or 3-brane as it is often called, is embedded in a five-dimensional (5D) bulk spacetime.
Depending on the functional expression of the model's warp factor, the 3-brane is characterized as thin (e.g. RS model) or thick. For introductory resources on braneworld models, the reader is directed to Refs. \cite{Maartens:2003tw, Brax:2004xh, Csaki:2004ay, Mannheim:2005br, Kribs:2006mq, Dzhunushaliev:2009va}.

In 1999, Chamblin, Hawking, and Reall (CHR)~\cite{Chamblin:1999by} pursued an intuitive approach in deriving the 5D geometry describing an analytic black hole (BH) localized on the 3-brane of an RS-II model (see also \footnote{In \cite{Dunajski:2018xoa}, the authors construct embeddings for any spherically symmetric Lorentzian metric in $3 + 1$ dimensions as a hypersurface in $\mathbb{R}^{4,1}$}). 
Their method involved replacing the flat four-dimensional part of the RS-II line element with that of the Schwarzschild spacetime.
From the perspective of a brane observer, the induced geometry is that of the usual Schwarzschild black hole. 
However, from a five-dimensional point of view, the CHR approach failed in providing a truly localized 5D object leading instead to an unstable black string solution~\cite{Gregory:1993vy}, where the singularity has an infinite extent along the extra dimension. 
For a non-exhaustive list of works attempting to provide a resolution to the black-hole localization problem, as well as works dedicated to the study of numerical or other exotic solutions, the reader is referred to 
\cite{EHM, Dadhich, tidal, Bruni, Papanto, KT, Dadhich2, CasadioNew, EFK, Frolov, EGK, Tanaka, KOT, Charmousis, Kofinas, Shanka, Karasik, GGI, Fitzpatrick, CGKM, Cuadros, Zegers, AS, Yoshino, Heydari, Andrianov1, Andrianov2, Dai, Kleihaus, Ovalle, KPZ, Harko, daRocha1, daRocha2, KPP, Banerjee, Chakraborty1, Chakraborty2, Chakraborty3, Nakas:2022hyp, Kudoh1, Kudoh2, Tanahashi, Figueras1, Page, Gubser, Wiseman, Sorkin1, Kudoh3, Sorkin2, Kleihaus2, Headrick, Figueras2, Kalisch, Emparan2, KNP1, KNP2, Rezvanjou, KNP3} and references therein.

As it was only recently pointed out by Nakas and Kanti (NK)~\cite{NK1,NK2}, the construction of a truly brane-localized 5D black hole in the RS-II model, requires a radically different approach to that of CHR.
In this letter, we extend the NK method by providing a General Embedding Algorithm (GEA) that allows~\emph{the consistent embedding of any static and spherically symmetric 4D geometry, in any single-brane 5D braneworld model with an injective and nonsingular warp factor}.
Additionally, we complement the GEA with a method that enables one to derive the field theory necessary to support the complete 5D braneworld geometry.
The herein proposed field-theory reconstruction technique provides \emph{a unified description for all braneworld models that exhibit the same brane-induced geometry}.

Upon utilizing the aforementioned methods, we present, for the first time, a field theory that is able to support the five-dimensional braneworld extension of the Schwarzschild geometry for any warp factor, and provide the consistent embedding of a de Sitter brane in any 5D braneworld model.
Finally, we delve into the multitude of possible  future applications of our method.

\noindent\textbf{Conventions and notation:} We adopt the $(-,+,+,+,+)$ signature for the metric and use natural units. Capital Latin letters will be employed to represent indices within the 5D bulk spacetime, while lowercase Greek letters will be used to denote indices related to the 4D brane. Note also that  $(\pa\phi)^2\equiv g^{MN}\pa_M\phi\, \pa_N\phi$. Throughout the manuscript, abiding by the established terminology in the community, we refer to the Weyl rescaling of the metric as conformal transformation.


\SectionStyle{General Embedding Algorithm (GEA)}\sectionbookmark{General Embedding Algorithm (GEA)}---The herein proposed GEA can be formulated in a number of straightforward steps, which are thoroughly presented below:

$\bullet\,\,$ \textbf{\emph{Step 0}:}  Our starting point is the line-element
\beq\label{eq:st0}
ds^2=e^{2A(y)} \left(-dt^2+dr^2+r^2d \Omega^2_2\right)+dy^2\,,
\eeq
which describes a flat 3-brane in a warped 5D spacetime, with $y \in \mathbb{R}$ being the extra dimension and the brane located at $y=0$.
The warping effect is encoded in the \emph{warp factor} $e^{2A(y)}$, while $d \Omega^2_2=d\vartheta^2+\sin^2\vartheta\, d\varphi^2$ is the line element of the unit 2-sphere. 
To simplify the formalism and without loss of generality, we assume a $\mathbf{Z}_2$-symmetric bulk so that the values of the warp factor for $y\leq0$ are uniquely mapped to those in the region $y\geq 0$. The extension to $\mathbf{Z}_2$-asymmetric models (e.g.~\cite{Takahashi:2004ss}) is trivial.

\hypertarget{st1}{$\bullet\,\,$ \textbf{\emph{Step 1}:}} The first step is to recast the 5D spacetime in a conformally-flat form. 
To this end, assuming an injective (one-to-one) and nonsingular warp factor, we apply the \textit{invertible} coordinate transformation $dy^2=e^{2A(y)} dz^2$ and we are led to 
\eq$\label{eq:z-y}
z(y)\equiv \int_0^{y} dw\, e^{-A(w)}\,.$
The above definition, maps $y=0$ to $z=0$ and thus ensures that the value of the warp factor on the brane does not change after the introduction of the new bulk coordinate $z$.
The line element \eqref{eq:st0} is now brought to the conformally-flat form
\eq$\label{eq:con-flat}
ds^2=e^{2A(y(z))} \left(-dt^2+dr^2+r^2d \Omega^2_2+dz^2\right)\,,$
where the warp factor multiples a flat 5D spacetime.

\hypertarget{st2}{$\bullet\,\,$ \textbf{\emph{Step 2}:}} Next, we express the flat component of the line element~\eqref{eq:con-flat} in five-dimensional spherical coordinates by performing the transformation
\eq$\label{eq:5D-sph-sym}
\{r,z\} \to \{\rho \sin \chi,\rho \cos \chi\}\,,\quad \chi \in [0,\pi]\,,$
with the inverse being
\beq
\{\rho,\chi\} \to \left\{\sqrt{r^2+z^2},\ \tan^{-1}\left(r/z\right)\right\}\,.
\label{eq:Inverse}
\eeq
Notice that the new radial coordinate $\rho$ is always positive-definite.
For more details about the geometrical set-up of the 5D spacetime in $\{t,\rho,\chi,\vartheta,\varphi\}$ coordinates, see \cite{NK2}.

\hypertarget{st3}{$\bullet\,\,$ \textbf{\emph{Step 3}:}} Upon promoting the $tt$ and $\rho\rho$ components of the conformally-flat metric to arbitrary functions of $\rho$, the five-dimensional line element reads
\eq$\label{eq:ds2_5d}
ds^2=e^{2A(y(\rho\cos\chi))} \left(-e^{\xi(\rho)}dt^2+e^{\eta(\rho)}d\rho^2+\rho^2d \Omega^2_3\right)\,,$
with $d\Omega_3^2=d\chi^2+\sin^2\chi\,d\vartheta^2+\sin^2\chi\sin^2\vartheta\,d\varphi^2$. The projection on the brane, is obtained by setting $\chi=\pi/2$ in~\eqref{eq:ds2_5d}, and the induced metric takes the form
\beq
ds_4^2=-e^{\xi(r)}dt^2+e^{\eta(r)}dr^2+r^2d\Omega^2_2\,.
\label{eq:ds2_4d}
\eeq
In the above, we have adopted the usual convention that the warp factor is normalized to unity on the brane, thus, $A(0)=0$.
The metric functions $\xi$ and $\eta$ retain their functional forms under $\rho \to r$. Hence, \textit{any 4D static and spherically symmetric geometry can be consistently embedded in any 5D braneworld model, characterized by an injective and nonsingular warp factor}.
Note also that in contrast to the CHR approach~\cite{Chamblin:1999by}, \textit{the 
preceding set-up ensures that any brane singularity, if present, remains confined on the brane}.
This is a consequence of the manifestation of spacetime singularities via the divergence of the curvature invariant quantities at $\rho=0$, which, by definition~\eqref{eq:Inverse}, vanishes only when both $r=0$ and $z=y=0$ are satisfied.

\hypertarget{st4}{$\bullet\,\,$ \textbf{\emph{Step 4}:}} As a final step, one may express the line element \eqref{eq:ds2_5d} in the original coordinate system $\{t,r,\vartheta,\varphi,y\}$ by following the inverse transformations~\eqref{eq:Inverse} to find
\bea
\label{eq:Gen_5D_ds2_r_y}
ds^2&=e^{2A(y)}\left\{-e^{\xi(r,y)} dt^2+\frac{dr^2}{r^2+z^2(y)}\biggl[e^{\eta(r,y)}r^2+z^2(y)\biggr]\right.\nonum\\[1mm]
&\hspace{4em}+r^2d\Omega_2^2+\left.\frac{2r z(y)\,e^{-A(y)}\left[e^{\eta(r,y)}-1 \right]}{r^2+z^2(y)}drdy \right\}\nonum\\[1mm]
&\hspace*{-7em}+\frac{dy^2}{r^2+z^2(y)}\left[r^2+e^{\eta(r,y)}z^2(y)\right]\,,\\[-0.3mm]\nonum
\eea
where $\xi(r,y)\equiv\xi\big(\sqrt{r^2+z^2(y)}\big)$, $\eta(r,y)\equiv\eta\big(\sqrt{r^2+z^2(y)}\big)$ and $z(y)$ is given by \eqref{eq:z-y}.
This coordinate system is considerably more convenient for the examination of the junction conditions on the brane and the computation of the brane matter content.


\SectionStyle{Field-theory reconstruction}\sectionbookmark{Field-theory reconstruction}---To provide a complete answer to the brane-localization problem, it is necessary to not only determine the appropriate 5D geometry, but also supplement it with the bulk field-theory content. To this end, the question that we are now called to answer is: \textit{Which field theory can support the 5D geometry and how can we reconstruct it?}
In what follows, \textit{we prove that the form of the required field theory capable of supporting a brane-localized geometry in a 5D braneworld scenario, depends solely on the brane configuration and not on the choice of the warp factor.
The difference in the warping of the extra dimension is encoded in the profile of a scalar field, but not in the field theory per se}.

Our field-theory reconstruction method stems from the simple observation that the line element \eqref{eq:ds2_5d} can be written in the form
\eq$\label{eq:ftr1}
ds^2= g_{MN}\,dx^Mdx^N=\psi^2\,\tilde{g}_{MN}\,dx^Mdx^N=\psi^2 d\tilde{s}^2\,,$
with $\psi\equiv e^{A(y(\rho\cos\chi))}$ and $g_{MN}$, $\tilde{g}_{MN}$ denoting the components of the warped and an auxiliary unwarped 5D geometry, respectively.
Upon interpreting the warp factor of the braneworld model as a conformal factor $\psi^2$, the two metrics become conformally related to one another. 
This allows us to establish a direct mapping/correspondence between the field theories associated with the two geometries via the well-known properties of the conformal transformation (see e.g.~\cite{Faraoni:1998qx, Dabrowski:2008kx}).
Consequently, it is convenient to initially derive the field theory that underlies the unwarped geometry $\tilde{g}_{MN}$, and then, by utilizing the conformal transformation~\eqref{eq:ftr1}, we effortlessly obtain the field theory that supports the warped metric.

Even though our method can be applied to any theory of gravity, for illustration purposes and with subsequent applications in mind, let us consider a particular example.
Supposing that the action functional \eqref{eq:uw-act-gen} admits the unwarped metric $\tilde{g}_{MN}$ in its spectrum of solutions, it is straightforward to deduce that the conformal transformation \eqref{eq:ftr1} leads us to the action~\eqref{eq:w-act-gen} associated with the warped metric $g_{MN}$. 
\begin{widetext}

\eq$\label{eq:uw-act-gen}
S=\frac{1}{\kappa_5}\int d^5x \sqrt{-\tilde{g}}\left[f(\phi)\tilde{R}+h(\phi) \tilde{\mathcal{G}}-\omega(\phi)(\tilde{\pa}\phi)^2-V(\phi)-\mathcal{L}_X\left(\tilde{g}_{MN},\widetilde{\text{other fields}}, \widetilde{\text{other h.o.c.}}\right) \right]\,,$
{\fontsize{8}{8}\selectfont
\bal$\label{eq:w-act-gen}
S=\frac{1}{\kappa_5}\int d^5x \sqrt{-g}\,\Bigg\{&\frac{f(\phi)}{\psi^3}\Bigg[R+8\frac{\square\psi}{\psi}-\frac{20}{\psi^2}(\pa\psi)^2\Bigg]
+\frac{h(\phi)}{\psi}\Bigg[\mathcal{G}-\frac{8}{\psi}\Big(2 R_{MN}\nabla^M\nabla^N \psi -R\, \square \psi\Big)
+\frac{12}{\psi^2}\Big( 2(\square\psi)^2-2\nabla_M\nabla_N\psi \nabla^M\nabla^N\psi -R(\pa \psi)^2 \Big)\nonum\\[1mm]
&-\frac{24}{\psi^3}\Bigg(4(\square\psi)(\pa\psi)^2-\frac{5}{\psi}(\pa_L\psi \pa^L\psi)^2 \Bigg)\Bigg]-\frac{\omega(\phi)}{\psi^3}(\pa\phi)^2 -\frac{V(\phi)}{\psi^5}-\frac{1}{\psi^5}\,\mathcal{L}_X\left(\frac{g_{MN}}{\psi^2},\text{other fields, other h.o.c.}\right)\Bigg\}\,.$}


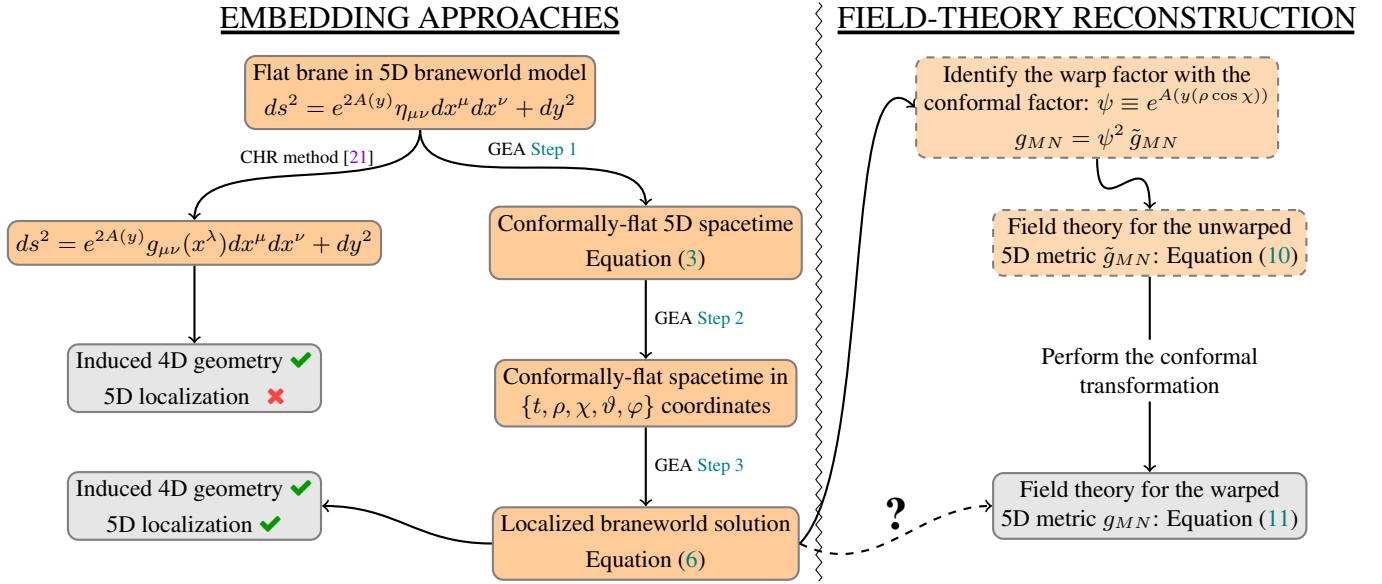
\begin{figure*}
\centering
\begin{tikzpicture}[auto, scale=1, every node/.style={scale=1}]

\tikzstyle{place}=[rectangle,rounded corners,draw=black!50,fill=orange!40,thick]
\tikzstyle{place2}=[rectangle,rounded corners,draw=black!50,dashed,fill=orange!30,thick]
\tikzstyle{result}=[rectangle,rounded corners,draw=black!50,fill=gray!20,thick]
\tikzstyle{pre}=[<-,shorten <=1pt,>=stealth’,semithick]
\tikzstyle{post}=[->,shorten >=1pt,>=stealth’,semithick]

\node[align=center] (GEA) at (0,1) [rectangle,draw=white,fill=white] {\large{EMBEDDING APPROACHES}};
\draw[black,line width=0.5mm,-] (-2.65,0.8) -- (2.65,0.8);

\node[align=center] (FTR) at (9,1) [rectangle,draw=white,fill=white] {\large{FIELD-THEORY RECONSTRUCTION}};
\draw[black,line width=0.5mm,-] (5.55,0.8) -- (12.42,0.8);

\draw[decoration = {zigzag,segment length = 2mm, amplitude = 0.4mm},decorate] (5.3,1.2) -- (5.3,-6.5);

\node[align=center] (CHR1) at (-3,-2) [place] {$ds^2=e^{2A(y)}g_{\mu\nu}(x^\lam)dx^\mu dx^\nu+dy^2$};
\node[align=center] (CHR2) at (-3,-3.8) [result] {Induced 4D geometry \textcolor{nicegreen}{\faCheck}\\[1mm] 5D localization \hspace{0.2em} \textcolor{nicegred}{\faTimes}};

\node[align=center] (S0) at (0,0) [place] {Flat brane in 5D braneworld model\\[1mm]
$ds^2=e^{2A(y)} \eta_{\mu\nu}dx^{\mu}dx^{\nu}+dy^2$};
\node[align=center] (S1) at (3,-2) [place] {Conformally-flat 5D spacetime\\[1mm]
Equation \eqref{eq:con-flat}};
\node[align=center] (S2) at (3,-4) [place] {Conformally-flat spacetime in\\[0.1mm] $\{t,\rho,\chi,\vartheta,\varphi\}$ coordinates};
\node[align=center] (S3) at (3,-6) [place] {Localized braneworld solution\\[1mm] Equation \eqref{eq:ds2_5d}};
\node[align=center] (S4) at (-3,-5.5) [result] {Induced 4D geometry \textcolor{nicegreen}{\faCheck}\\[1mm] 5D localization \textcolor{nicegreen}{\faCheck}};

\node[align=center] (FTR0) at (9,-0.2) [place2] {Identify the warp factor with the\\[0.1mm]
conformal factor: $\psi\equiv e^{A(y(\rho\cos\chi))}$\\[1mm]
$g_{MN}=\psi^2\, \tilde{g}_{MN}$};
\node[align=center] (FTR1) at (9.7,-2) [place2] {Field theory for the unwarped\\[0.1mm] 5D metric $\tilde{g}_{MN}$: Equation \eqref{eq:uw-act-gen}};
\node[align=center] (con) at (9.7,-3.7) [rectangle,draw=white,fill=white] {Perform the conformal\\[0.1mm] transformation};
\node[align=center] (FTR2) at (9.7,-5.5) [result] {Field theory for the warped\\[0.1mm]
5D metric $g_{MN}$: Equation \eqref{eq:w-act-gen}};

\draw[thick,black,->] (S0.south) .. controls +(0,-1) and +(0,1) .. (CHR1.north) node[draw=none,fill=none,font=\scriptsize,midway,above] {CHR method \cite{Chamblin:1999by}};
\draw[thick,black,->] (CHR1.south) -- (CHR2.north);

\draw[thick,black,->] (S0.south) .. controls +(0,-1) and +(0,1) .. (S1.north)
node[draw=none,fill=none,font=\scriptsize,midway,above] {GEA \hyperlink{st1}{Step 1}};
\draw[thick,black,->] (S1.south) -- node[draw=none,fill=none,font=\scriptsize,midway] {GEA \hyperlink{st2}{Step 2}} (S2.north);
\draw[thick,black,->] (S2.south) -- node[draw=none,fill=none,font=\scriptsize,midway] {GEA \hyperlink{st3}{Step 3}} (S3.north);
\draw[thick,black,->] (S3.west) .. controls +(-1,0) and +(1,0) .. (S4.east);

\draw[thick,black,->] (S3.east) .. controls +(1,1) and +(-1,1) .. (FTR0.west);
\draw[thick,black,->] (FTR0.south) .. controls +(0,-1) and +(0,1) .. (FTR1.north);
\draw[thick,black,-] (FTR1.south) -- (con.north);
\draw[thick,black,->] (con.south) -- (FTR2.north);
\draw[thick,black,dashed,->] (S3.east) .. controls +(1,-1/2) and +(-1,0) .. (FTR2.west)
node[draw=none,fill=none,midway,above] {\textbf{\LARGE{?}}};

\end{tikzpicture}
\caption{Schematic representation of our General Embedding Algorithm (GEA) compared to the Champlin, Hawking, and Reall's (CHR) approach.}
\label{Fig: algorithm}
\end{figure*}


\end{widetext}
In the above relations, $R$ is the 5D Ricci scalar, $\mathcal{G}\equiv R^2-4 R^{MN} R_{MN}+R^{KLMN} R_{KLMN}$ is the Gauss-Bonnet curvature invariant, $\phi$ is a real scalar field, and $f(\phi)$, $h(\phi)$, $\omega(\phi)$, $V(\phi)$ are arbitrary model functions to be determined via the field equations.
In addition, the constant $\kappa_5$ is directly related to the fundamental five-dimensional Planck scale $M_{Pl(5)}$, while the general Lagrangian density $\mathcal{L}_X$ incorporates a plethora of models including even additional higher-order curvature (h.o.c.) terms.
For ease of reference, we provide a concise overview of our complete method in the flowchart of Fig.~\ref{Fig: algorithm}.

Note also here that with the 5D bulk action \eqref{eq:w-act-gen} at hand, which supports the warped metric $g_{MN}$, the analysis of junction conditions on the brane is straightforward. 
This, in turn, facilitates the derivation of the effective theory on the four-dimensional 3-brane (see e.g. \cite{Shiromizu:1999wj}). For applications of this analysis to black-hole solutions in the context of the RS-II model, refer to \cite{NK1,NK2}.

To illustrate the versatility of our method, the following two sections are devoted to the field-theory reconstruction regarding the 5D braneworld extensions for two of the simplest curved 4D geometries, namely the Schwarzschild and the de Sitter (dS).


\SectionStyle{Localized braneworld black holes}\sectionbookmark{Localized braneworld black holes}---In the context of the RS-II braneworld model~\cite{Randall:1999vf}, the exact geometry describing a five-dimensional BH that is exponentially localized close to the brane and induces the Schwarzschild metric on the brane, has been obtained in~\cite{NK1}.
In this section, we provide for the first time an answer to the long-standing question regarding the bulk field theory necessary to support a braneworld BH with the aforementioned features.
Most importantly, by virtue of the herein introduced GEA we can tackle the above problem in its full generality without restricting our analysis to the RS-II model.

In terms of the spherical coordinates of~\hyperlink{st2}{Step 2}, the five-dimensional extension of the Schwarzschild geometry in an arbitrary braneworld scenario, according to ~\hyperlink{st3}{Step 3}, is expressed as
\eq$\label{eq:loc_BH_metric}
ds^2=e^{2A(y(\rho\cos\chi))} \big(-e^{\xi(\rho)}\,dt^2+e^{-\xi(\rho)}d\rho^2+\rho^2\,d\Omega_3^2\big),$
with $e^{\xi(\rho)}=1-2M/\rho$ and $M$ being the black-hole mass parameter.
Following along the lines of the analysis performed in \cite{NK1} for the RS-II model, one can verify that for any nonsingular warp factor, the curvature invariant quantities emanating from the line-element \eqref{eq:loc_BH_metric}, exhibit a curvature singularity only at $\rho=0$, where both $r=0$ and $z=y=0$.
This means that the black-hole singularity resides strictly on the brane, while the geometrical structure of the black-hole horizon is dictated by the equation $e^{\xi(\rho)}=0 \Rightarrow r^2+z^2(y)=4M^2$ ensuing from radial null-trajectories in the 5D spacetime \eqref{eq:loc_BH_metric}.

In order to facilitate the reconstruction of the field theory, we first consider the unwarped geometry associated with the line element~\eqref{eq:loc_BH_metric}, namely
\beq\label{eq:un_metr}
d\tilde{s}^2=\tilde{g}_{MN}dx^M dx^N=-e^{\xi(\rho)}\,dt^2+e^{-\xi(\rho)}d\rho^2+\rho^2\,d\Omega_3^2\,,
\eeq
for which the corresponding components of the Einstein tensor assume the particularly simple form
\beq
\tilde{G}^M_{\,\,\,N}=\text{diag}\left[-3M/\rho^3,-3M/\rho^3,0,0,0\right]\,.
\label{eq:GMN_BH}
\eeq
Next, we consider the general scalar-tensor theory \eqref{eq:uw-act-gen}, with $\phi=\phi(\rho)$ and $\mathcal{L}_X=0$.
The field equations arising from this action functional consist of the equation of motion (EOM) for $\phi$ and the bulk Einstein equations. 
As is typically the case with scalar-tensor theories, the scalar-field EOM can be derived from the Einstein EOM under an appropriate manipulation. 
In our case, we have three independent equations in total, which can be taken to be the $({}^{t}_{\,\, t})$, $({}^{t}_{\,\, t})-({}^{\rho}_{\,\, \rho})$, and $({}^{t}_{\,\, t})-({}^{\chi}_{\,\, \chi})$ Einstein equations. However, since the field theory contains five unknown functions (four model functions and the scalar field), it is natural to anticipate that this system is sufficiently general to accommodate the unwarped geometry \eqref{eq:un_metr} as a solution.
In order to close the system of equations, one has to assume the functional form for two out of the five free functions.
To this end, by assuming a specific expression for the scalar function $\phi(\rho)$ and the non-minimal coupling function $f(\phi)$, we are led to the following expressions for the functions of the model
\eq$
\begin{gathered}
\phi(\rho)=\frac{M}{\rho^3}\,,\quad f(\phi)=48 \alp \phi^{\frac{1}{3}}\,,\quad h(\phi)=\frac{\alp}{\phi^{\frac{2}{3}}}\,,\\[1mm]
\omega(\phi)=-\frac{16 \alp}{\phi^{\frac{5}{3}}}\,,\quad V(\phi)=-48 \alp \left[\phi^{\frac{4}{3}}-3\frac{\phi}{M^{\frac{2}{3}}} \right]\,,
\end{gathered}
\label{eq:uw-bh-act}$
where $\alp$ is a constant with units (\emph{length})${}^{2/3}$.

Notice that although our assumptions led us to a simple bulk field theory, the black-hole mass parameter $M$ appears explicitly in the scalar potential $V(\phi)$, and as such, it assumes the role of a coupling constant in our theory.
Consequently, after reintroducing the warp factor as a conformal transformation into \eqref{eq:un_metr}, the resulting theory will describe an isolated eternal braneworld BH of fixed mass, a feature that is troubling.
A truly satisfactory reconstructed theory would not have the mass parameter $M$ of the black hole to appear as a coupling constant in the Lagrangian.
In our case though, the field theory \eqref{eq:uw-bh-act} is only a particular example of the general theory \eqref{eq:uw-act-gen} with $\mathcal{L}_X=0$. A different, inspired assumption for two of the free functions may result to the desired theory. Such an effort however requires a trial-and-error approach and goes beyond the scope of this section that aims to serve as an illustration of the method.
It is also important to emphasize at this point, that the preceding issue plagues the vast majority of field-theory reconstruction attempts performed in the literature, where attention is placed on the features of the overall geometry but not on the model functions of the theory.

Reintroducing now the warp factor as a conformal transformation of the unwarped metric, according to \eqref{eq:ftr1}, and utilizing the well-known conformal transformation rules (see e.g.~\cite{Faraoni:1998qx, Dabrowski:2008kx}), one can directly map the action~\eqref{eq:uw-act-gen} (with $\mathcal{L}_X=0$), to the conformally-transformed frame \eqref{eq:w-act-gen} (with $\mathcal{L}_X=0$) that supports the warped spacetime~\eqref{eq:loc_BH_metric}.
Note that the field equations regarding the theory of the warped metric are guaranteed to be free of Ostrogradsky instabilities and ghosts.
This confident assertion is founded on two known facts.
On the one hand, the unwarped Lagrangian density $\mathcal{L}=f(\phi)\tilde{R}+h(\phi)\tilde{\mathcal{G}}-\omega(\phi)(\tilde{\pa}\phi)^2-V(\phi)$ constitutes a special case of the Horndeski theory (see e.g. \cite{Kobayashi:2019hrl}), while on the other hand, the conformally-transformed/warped theory remains also within the Horndeski class (see \cite{Bettoni:2013diz}).

Finally, by following \hyperlink{st4}{Step 4}, the line element of the brane\-world extension of the Schwarzschild black hole \eqref{eq:loc_BH_metric} can be expressed in the coordinate system $\{t,r,\vartheta,\varphi,y\}$ and brought to the form of \eqref{eq:Gen_5D_ds2_r_y}, with $e^{\xi(r,y)}=e^{-\eta(r,y)}=1-\frac{2M}{\sqrt{r^2+z^2(y)}}$.
In this form and upon choosing the warp factor of the model, the analysis regarding the junction conditions on the brane as well as the energy conditions of the stress-energy tensor, along with its induced effects on the brane, can be performed more efficiently.
In the context of the RS-II model, where $A(y)=-k|y|$ and $z(y)=sgn(y)\,(e^{k |y|}-1)/k$, and despite the absence of the knowledge of a bulk field theory that supports the geometry, the aforementioned analysis has been performed in~\cite{NK1} in a theory-agnostic way by utilizing the Einstein equations in combination with the known components of the Einstein tensor. 
As the analysis therein revealed, the bulk stress-energy tensor emerging from any theory with Einstein equations of the form $G_{MN}=T_{MN}$, satisfies the energy conditions on the brane, while a local violation takes place in the bulk, which is necessary for the BH localization close to the 3-brane. This violation can potentially be avoided entirely with an appropriate choice for the warp factor.


\SectionStyle{De Sitter brane in a 5D bulk}\sectionbookmark{De Sitter brane in a 5D bulk}---As a second application of our GEA, we consider the embedding of a $dS_4$ brane in a five-dimensional bulk. Starting from the induced metric on the brane that is here identified with the 4D de Sitter geometry, we introduce, according to~\hyperlink{st3}{Step 3}, the following auxiliary unwarped 5D geometry
\beq
d\tilde{s}^2=-\left(1-\frac{\Lambda_{\rm eff}}{3}\rho^2\right)dt^2+\frac{d\rho^2}{\left(1-\frac{\Lambda_{\rm eff}}{3}\rho^2\right)}+\rho^2d\Omega_3^2\,,
\label{eq:Unwarped_(A)dS}
\eeq
where $\Lambda_{\rm eff}>0$ is the effective cosmological constant on the brane. 
The field theory supporting~\eqref{eq:Unwarped_(A)dS} is readily obtained and reads
\beq
S=\frac{1}{\kappa_5}\int d^5x \sqrt{-\tilde{g}}\left(\frac{\tilde{R}}{2}-2\Lambda_{\rm eff}\right)\,.
\label{eq:aux_5D_dS_action}
\eeq
Having specified both the geometry~\eqref{eq:Unwarped_(A)dS} and the theory~\eqref{eq:aux_5D_dS_action} in the auxiliary unwarped frame, we can now easily obtain the warped 5D geometry along with its corresponding field theory. By performing the conformal transformation~\eqref{eq:ftr1}, the auxiliary action~\eqref{eq:aux_5D_dS_action} is straightforwardly mapped to the bulk field theory

\vspace*{-1em}
{\fontsize{9}{9}\selectfont
\beq
S=\int d^5x \frac{\sqrt{-g}}{\kappa_5}\,\Bigg\{\frac{1}{2\psi^3}\Bigg[R+8\frac{\square\psi}{\psi}-\frac{20}{\psi^2}(\pa\psi)^2\Bigg]-\frac{2\Lambda_{\rm eff}}{\psi^5}\Bigg\}\,.
\label{eq:5D_dS_action_warped}
\eeq}
\hspace*{-0.3em}The field equations originating from~\eqref{eq:5D_dS_action_warped}, admit as a solution the warped 5D geometry~\eqref{eq:ds2_5d} with metric functions $e^{\xi(\rho)}=e^{-\eta(\rho)}=1-\Lambda_{\text{eff}}\, \rho^2/3$, or equivalently the geometry~\eqref{eq:Gen_5D_ds2_r_y} with metric functions
\beq
e^{\xi(r,y)}=e^{-\eta(r,y)}=1-\frac{\Lambda_{\rm eff}}{3}\left[r^2+z^2(y)\right]\,.
\label{eq:dS_metric_funcs}
\eeq
Consequently, the line element~(\ref{eq:Gen_5D_ds2_r_y},\,\ref{eq:dS_metric_funcs}) describes the consistent embedding of a $dS_4$ brane in an arbitrary braneworld bulk spacetime characterized by an injective warp function $A(y)$. 
For thick braneworld models, field theories that support embeddings of a de Sitter brane in an $AdS_5$ bulk are known (see e.g.~\cite{Dzhunushaliev:2009va}), however, in these models, the warp function is not pre-defined but is rather determined by the theory's field equations. 
To our knowledge, the action functional~\eqref{eq:5D_dS_action_warped} with $\psi=e^{-k|y|}$, provides the first example in the literature of a field theory supporting the embedding of a $dS_4$ brane in the context of the RS-II model. 

It is also worth noting that any braneworld model of the form (\ref{eq:Gen_5D_ds2_r_y},\,\ref{eq:dS_metric_funcs}) that exhibits an $AdS_5$ asymptotic behavior, will adhere to the relation
\eq$\label{eq:R-asym}
\lim_{y\rightarrow \pm \infty} R=\lim_{\rho\rightarrow \infty} R=\frac{10 \Lambda_5}{3}\,,$
with $\Lambda_5<0$ being the bulk cosmological constant.
Hence, Eq. \eqref{eq:R-asym} will always relate the bulk cosmological constant $\Lambda_5$ with the induced cosmological constant $\Lambda_{\text{eff}}$ on the brane.
Especially for the RS-II model and in the $\{t,\rho,\chi,\vartheta,\varphi\}$ coordinate system, in which $A(y(\rho \cos\chi))=-\ln(1+k\rho |\cos\chi|)$, the Ricci scalar in the bulk is given by
\eq$\label{eq:R-dS-RS2}
R=20\left(\frac{\Lambda_{\rm eff}}{3}-k^2 \right)\,.$
Therefore, for this case, $\Lambda_5=2\left(\Lambda_{\text{eff}}-3k^2\right)$, which entails the constraint $\Lambda_{\text{eff}}<3k^2$.

As a second example, we consider the commonly used thick-brane model characterized by the warp function $A(y)=-\ln{\cosh (\sigma y)}$ (see \cite{Sui:2020atb,Li:2022kly,Moreira:2021cta,Moreira:2022zmx,Lima}),
where $\sigma$ is a parameter associated with the brane thickness. 
In the 5D spherical coordinates, the warp function is expressed as $A(y(\rho \cos\chi))=-\frac{1}{2}\ln(1+\sigma^2 \rho^2\cos^2\chi)$, while the bulk cosmological constant is evaluated to be $\Lambda_5=-\left(30 \sigma^2+4\Leff\right)/5$. Thus, independently of the value of $\sigma$, $\Lambda_5<0$ for any $\Leff>0$.

It is worth mentioning that the application of the CHR embedding approach in the case of a $dS_4$ brane geometry and for the RS-II model, leads to a bulk spacetime which is characterized by the Ricci scalar
\beq
R=4\left(e^{2k|y|}\Leff-5k^2 \right)\,.
\eeq
The above curvature invariant quantity diverges in the limit $|y|\to \infty$ which is to be contrasted with~\eqref{eq:R-dS-RS2} that is free of any pathology. 
This example serves as additional evidence for the effectiveness of our General Embedding Algorithm in generating genuinely five-dimensional brane-localized geometries.


\SectionStyle{Summary and future directions of research}\sectionbookmark{Summary and future directions of research}--- In this letter, we have introduced a General Embedding Algorithm that allows one to consistently extend any four-dimensional static and spherically symmetric geometry into any five-dimensional single-brane braneworld model with a nonsingular and injective warp factor. In contrast to previous embedding approaches, our GEA, has the important advantage of ensuring, by construction, the localization of the bulk gravity close to the brane. As a consequence, the resultant 5D geometries are devoid of undesired features such as exhibiting a black-string topology or possessing curvature invariants that diverge asymptotically. 
We also supplemented the GEA by proposing a general method that greatly assists the reconstruction of the bulk field theory supporting the 5D braneworld geometries. 
Consequently, in effect, \emph{we provide a complete procedure for the consistent embedding of 4D geometries into braneworld models}. 
In light of the results and insights presented in this work, there are several promising avenues for future research.
In what follows, we briefly outline some of the main directions that merit further investigation.

Firstly, due to the fact that our method has been formulated in the most general way, one is presented with the opportunity to explore consistent embeddings of other interesting static and spherically symmetric spacetimes that go beyond the black-hole and de Sitter cases presented here. 
For example, one may consider embeddings of 4D wormhole configurations, regular black holes \cite{Neves:2021dqx}, models of stars, cosmological \cite{Binetruy:1999ut,Langlois:2001dy,Bogdanos:2006pf,Antonini:2019qkt, Antonini:2021xar}, or other kinds of solutions \cite{Bazeia:2023czl}. 
Furthermore, given the connections established between the brane, the bulk and the auxiliary unwarped domains, one may follow a different approach to the one followed in this work. 
That is, instead of a ``brane-first" approach, one may begin by considering field theories in the unwarped frame and then obtain the corresponding brane and bulk geometries in the warped frame. In this way, novel geometries with potentially interesting characteristics could emerge.
Secondly, the adaptation and extension of our GEA to accommodate axially-symmetric spacetimes is yet another research direction that one could pursue.
This, although not a trivial task, has great relevance for astrophysically-realistic configurations that exhibit some amount of rotation.

As we have already seen in the corresponding section, our brane\-world black-hole solutions, induce the Schwarzschild geometry on the brane independently of the choice of the warp factor. However, perturbative effects are expected to give rise to nontrivial signatures on the brane associated with the bulk structure of these objects. 
Hence, a detailed stability analysis for these configurations, remains an open question of critical importance. 
Given the freedom in choosing the warp factor, a first step in this direction could be the systematic classification of braneworld models with respect to the stability of their BH solutions. 
Additionally, this freedom presents yet another intriguing possibility, by potentially allowing configurations that are unstable in four dimensions to become stable once embedded in the appropriate braneworld scenario.

Lastly, braneworld models with $AdS_5$ asymptotics and a nonflat brane-induced geometry could also be considered in the context of holography \cite{Maldacena:1997re, Gubser:1998bc, Witten:1998qj,Antonini:2019qkt, Antonini:2021xar,Geng:2021mic,Geng:2022dua,Geng:2023iqd} once they undergo the appropriate uplift. 
In this regard, they can be used to describe various field-theory phenomena, including chiral symmetry breaking \cite{DaRold:2005mxj,Alho:2013dka}, confinement/deconfinement \cite{BallonBayona:2007vp}, and other quantum processes \cite{Mori:2023swn}.

\SectionStyle{Acknowledgments}\sectionbookmark{Acknowledgments}---T.N. is thankful to Athanasios Bakopoulos for many enlightening discussions.
The research project was supported by the Hellenic Foundation for Research and Innovation (H.F.R.I.) under the ``3rd Call for H.F.R.I. Research Projects to support Post-Doctoral Researchers" (Project Number: 7212).
T.N. also acknowledges networking support through the program ``Mezin\'{a}rodn\'{i} mobilita v\'{y}zkumn\'{y}ch, technick\'{y}ch a administrativn\'{i}ch pracovn\'{i}k\r{u} v\'{y}zkumn\'{y}ch organizaci'' (Registra\v{c}n\'{i} \v{c}\'{i}slo projektu: CZ.02.2.69/0.0/0.0/18\_053/0017871) and is very grateful to the Research Centre for Theoretical Physics and Astrophysics of the Institute of Physics at the Silesian University in Opava for the hospitality during the course of this work.
T.D.P., and Z.S. acknowledge the support of the Research Centre for Theoretical Physics and Astrophysics of the Institute of Physics at the Silesian University in Opava. 

\bibliographystyle{apsrev4-1}
\bibliography{References.bib}

\end{document}